\documentclass[draft]{agujournal2019}
\usepackage{url} 
\usepackage[finalnew]{trackchanges} 
\usepackage{soul}
\draftfalse

\journalname{Geophysical Research Letters}
\begin{document}

\title{Soft X-ray imaging of magnetopause reconnection outflows under low plasma-$\beta$ solar wind conditions}

%
%

\authors{Yosuke Matsumoto\affil{1} and Yoshizumi Miyoshi \affil{2}
}
\affiliation{1}{Institute for Advanced Academic Research, Chiba University, Chiba, Japan}
\affiliation{2}{Institute for Space-Earth Environmental Research, Nagoya University, Nagoya, Japan}

\correspondingauthor{Yosuke Matsumoto}{ymatumot@chiba-u.jp}

\begin{keypoints}
\item Global magnetospheric MHD modeling of soft X-ray emission is examined under low-temperature solar wind conditions
\item Soft X-ray emission by the solar wind charge-exchange process is expected to be very bright around the dayside reconnection region
\item X-ray map can reflect plasma jets under cold solar wind allowing visualization of the reconnection region as observed in the solar corona
\end{keypoints}

\begin{abstract}
We examined soft X-ray emission by the solar wind charge-exchange process around the Earth's magnetosphere using a global magnetohydrodynamic simulation model. The dayside magnetopause reconnection heats and accelerates the plasma whereby the X-ray emission becomes as bright as $\sim 6 \times 10^{-6} {\rm\ eV}\ {\rm cm}^{-3}\ {\rm s}^{-1}$ under the southward interplanetary magnetic field conditions. In particular, under low plasma-$\beta$ solar wind conditions, we found that the X-ray intensity reflects the bulk motion of outflows from the reconnection region. We propose that this particular solar wind condition would allow visualization of the mesoscale magnetopause reconnection site, as observed in the solar corona.
\end{abstract}

\section*{Plain Language Summary}
A charge exchange between highly charged-state ions in the solar wind and neutral atoms is understood as a bright source of soft X-ray in space. It has been suggested that this emission helps visualize the global structures of the Earth's magnetosphere as a backlight; that is, we expect such an emission to be bright, in particular in the dayside solar wind (magnetosheath), and dark on the magnetosphere side. For validation and for an upcoming space telescope mission, we have developed a numerical model to provide the spatial distribution of the X-ray intensity. We conducted numerical simulations under various solar wind conditions. The model predicts that the X-ray emission is bright in the current layer near the reconnection region at the magnetospheric boundary. In particular, under low-temperature solar wind conditions, we found that the X-ray intensity reflects the bulk motion of reconnection jets, thus allowing visualization of a breaking of the terrestrial magnetic barrier.

\section{Introduction}
In-situ spacecraft observations have revealed the plasma dynamics in the Earth magnetosphere and the solar wind in response to variations in the activities on the Sun. Such observations provide ample opportunities to understand plasma kinetics, where energy release and dissipation by magnetic reconnection, collision-less shocks, and turbulence are of great interest. By contrast, global imaging of remote objects by observation of electromagnetic wave emissions, as in radio waves, optical light, infrared, X-, and gamma rays, is a common tool in astrophysics. Such remote imaging techniques have also been used for visualizing the near-Earth space environment (geospace). The plasmasphere in the inner magnetosphere was visualized by observing extreme ultraviolet emission \cite{nakamura_terrestrial_2000,burch_views_2001}. The detection of energetic neutral atoms generated through a charge-exchange process between protons and neutral (hydrogen) atoms has been a tool for providing global pictures of the ring current \cite{burch_views_2001,goldstein_five_2013} in the inner magnetosphere, the magnetosheath, and the cusp regions \cite{fuselier_energetic_2010,petrinec_neutral_2011}, as well as in heliospheric structures \cite{mccomas_global_2009}.

It has been suggested that observation of soft X-ray emissions could be useful as a remote imaging tool \cite{sibeck_imaging_2018}. The soft X-ray emission in this context was first recognized as an unknown source of X-ray enhancement discovered by the astrophysical X-ray space telescope, and was known as the long-term enhancement \cite{snowden_analysis_1994}. Such mysterious emissions were also found when the X-ray telescope observed the Hyakutake comet \cite{lisse_discovery_1996}. Later, it was found that the enhancement correlated well with solar wind proton flux variations \cite{cravens_temporal_2001}. We now understand that the emission is attributed to the charge exchange between highly charged-state heavy ions, such as ${\rm C^{6+}}$, ${\rm O^{7+}}$ or ${\rm O^{8+}}$ ions, in the solar wind and neutral atoms \cite{cravens_comet_1997}. This process is referred to as solar wind charge exchange (SWCX).

SWCX in the geospace was evidenced by spectrum-resolved X-ray observations by Chandra \cite{wargelin_chandra_2004}, XMM-Newton \cite{snowden_xmmnewton_2004,carter_identifying_2008,connor_exospheric_2019}, and Suzaku \cite{fujimoto_evidence_2007,ezoe_time_2010,ishikawa_suzaku_2013}. These observations with the spectral information in sub-keV energies revealed that enhanced counts at the energies expected for SWCX emission lines were observed when instruments pointed at regions with high densities of solar wind ions like the magnetosheath and the cusp.

After these successful observations, specially designed  missions of the X-ray space telescope, including SMILE (\url{https://www.cosmos.esa.int/web/smile/home}) and STORM (\url{https://stormmission.com/}), were proposed to visualize the magnetosphere through SWCX. Japanese GEOspace X-ray imager (GEO-X) project shares such scientific objectives and has been approved as a very small satellite mission. The GEO-X satellite is scheduled to be launched during the upcoming solar maximum and will be delivered to a low-latitude orbit at a distance of the lunar orbit \cite{ezoe_geo-x_2020}. 

Numerical modeling of the SWCX emission is necessary for mission design and to determine the scientific targets in advance. Such modeling also complements the observations to understand the magnetospheric dynamics behind them. Global magnetohydrodynamic (MHD) simulations of the Earth magnetosphere have been used to model X-ray emission in the dayside magnetosheath, the cusp \cite{kuntz_solar_2015,connor_exospheric_2019,connor_soft_2021}, and the low-latitude boundary layer subject to the Kelvin--Helmholtz vortex evolution \cite{sun_x-ray_2015}. While these simulation models provided intensity maps reflecting the shape of the magnetospheric boundaries, in this Letter, we propose that by employing global MHD simulations, the X-ray emission can provide unique information concerning the plasma dynamics around the magnetopause reconnection site under particular solar wind conditions. We expect to observe accelerating plasma outflows from the reconnection region  from the low-latitude orbit of GEO-X.

\section{Numerical Models}
We developed a global MHD simulation model of the magnetosphere by using the public MHD code CANS+, which adopts standard Godunov schemes, including the approximate Riemann solvers and the nonlinear interpolation schemes \cite{matsumoto_magnetohydrodynamic_2019}. In this study, we specifically used the Harten-Lax-van Leer (HLL) approximate Riemann solver \cite{hussaini_uniformly_1987} and the fifth-order, monotonicity-preserving (MP5) scheme \cite{suresh_accurate_1997}. We solved the modified MHD equations for numerically stable solutions with the dipole magnetic field by subtracting the potential field in the numerical flux calculation \cite{miyoshi_hlld_2010,guo_extended_2015}. 

The simulations were conducted in Cartesian coordinates with inner boundary conditions on a sphere surface at a radial distance of  $R=4\ R_{\rm E}$, where $R_{\rm E}$ is the Earth's radius. The numerical resolution in space was defined as $\Delta_X = \Delta_Y = \Delta_Z = 0.15\ R_{\rm E}$ in $X \le 54\ R_{\rm E}$ and $|Y|, |Z| \le 22.875\ R_{\rm E}$. The resolution gradually decreased in regions further outward and tailward. The overall domain covered $-30\ R_{\rm E} \le X \le 85.6\ R_{\rm E}$ and $-38.3\ R_{\rm E} \le Y, Z \le +38.3\ R_{\rm E}$ with $592 \times 355 \times 355$ computational cells. The solar wind plasma and the interplanetary magnetic field (IMF) were imposed as a boundary condition in the $Y$--$Z$ plane at $X=-30\ R_{\rm E}$. (Here the positive X- and Y-axes point to the tailward and dawnward directions, respectively.)

We calculated the X-ray emission intensity by the empirical model \cite{cravens_temporal_2001,connor_soft_2021} given as
\begin{equation}
I = \alpha N_{\rm p} N_{\rm H}\sqrt{v_{\rm th}^2+V} = \alpha N_{\rm p} N_{\rm H} \sqrt{\frac{3k_{\rm B} T}{M}+V^2} ~~ [{\rm eV\ cm^{-3}\ s^{-1}}],
\label{x-intensity}
\end{equation}
where $\alpha=6\times10^{-16} ~ {\rm eV\ cm^2}$ is the coefficient that incorporates information about the cross sections, the emission line energies, and the ion compositions in the solar wind,  $N_{\rm p}$ and $N_{\rm H}$ are the plasma and hydrogen number density, respectively, $v_{\rm th}$ is the plasma thermal speed for the plasma temperature $T$ with the Boltzmann constant $k_{\rm B}$ and the proton mass $M$, and $V$ is the bulk speed of plasma. The MHD model provides three-dimensional (3D) distributions of $N_{\rm p}$, $T$, $V$, whereas the hydrogen number density profile is given by a spherically symmetric model of the exosphere \cite{cravens_temporal_2001} as
\begin{equation}
\label{exo_model}
   N_{\rm H} = 25\left(\frac{10\ R_{\rm E}}{R}\right)^3 ~~ [{\rm cm^{-3}}] .
\end{equation}

Now we focus on the emission around the magnetopause reconnection site and consider the possibility of finding the plasma dynamics from the X-ray intensity map. For this purpose, we arrange eq. (\ref{x-intensity}) as
\begin{equation}
    I = \alpha N_{\rm p} N_{\rm H} \sqrt{\frac{3k_{\rm B} T}{M}+V^2} = \alpha N_{\rm p} N_{\rm H} V \sqrt{\frac{1.4}{M_{\rm s}^2}+1},
\label{x-intensity_mach}
\end{equation}
where the sonic Mach number $M_{\rm s}$ refers to the reconnection outflow, with the specific heat ratio of $5/3$. Using the relation between the outflow Mach number and the plasma $\beta$ in the inflow region \cite{soward_fast_1982,aurass_shock-excited_2002,seaton_analytical_2009}, we have
\begin{equation}
I = \alpha N_{\rm p} N_{\rm H} V\sqrt{\frac{1.4}{M_{\rm s}^2}+1} = \alpha N_{\rm p} N_{\rm H} V \sqrt{\frac{6+15\beta_{\rm sheath}}{10}+1} \label{x-intensity_beta},
\end{equation}
where $\beta_{\rm sheath}$ is the plasma $\beta$ defined in the magnetosheath (shock downstream). From eq. (\ref{x-intensity_beta}), we find that the X-ray intensity can reflect the plasma bulk motion under low-$\beta$ conditions in the shock downstream ($\beta_{\rm sheath}$) provided $N_{\rm H}$ is uniform within the scale of the current sheet. In the limit of $\beta_{\rm sheath}\ll1$, we expect X-ray emissions from the outflow with the strength of
\begin{equation}
    I = 1.0\times 10^{-5} \left(\frac{N_{\rm p}}{10~{\rm cm^{-3}}} \right)\left( \frac{10~R_{\rm E}}{R}\right)^3 \left(\frac{V}{500~{\rm km~s^{-1}}}\right) ~~ [{\rm eV\ cm^{-3}\ s^{-1}}].
\end{equation}
Next, we search for upstream solar wind conditions such that $\beta_{sheath}$ becomes less than unity. 

\begin{figure}
\includegraphics[width=\textwidth]{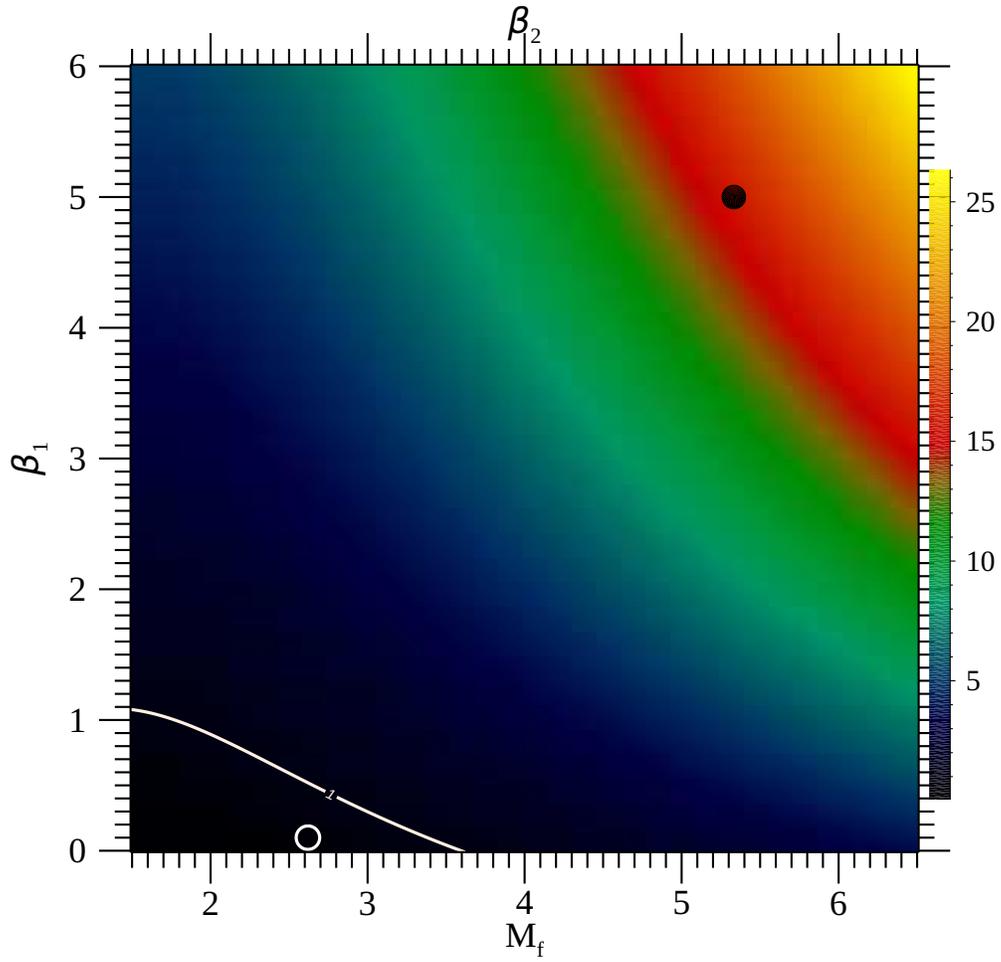}
\caption{Downstream plasma $\beta_2$ from the magnetosonic perpendicular shock jump condition as a function of the magnetosonic Mach number $M_{\rm f}$ and the upstream plasma $\beta_1$. The solid white line indicates $\beta_2=1$. Black filled and white open circles indicate the selected normal and low-$\beta$ solar wind conditions, respectively.} 
\label{RH}
\end{figure}

Figure \ref{RH} shows the downstream plasma $\beta$ calculated from the magnetosonic perpendicular shock jump condition. A typical solar wind parameter (black filled circle) results in a high $\beta$ ($\sim 15$) plasma in the magnetosheath. For these normal conditions, we expect the X-ray emission to reflect hot plasma distribution in the magnetosheath. By contrast, conditions in which the downstream $\beta$ becomes less than unity are limited to the left bottom corner of the diagram. In this study, we selected upstream parameters as indicated by the white open circle in Figure \ref{RH}, giving downstream $\beta \sim 0.5$, as a low-$\beta$ solar wind condition. Other upstream parameters are summarized in Table \ref{sw_params}.

\begin{table}
\caption{Upstream solar wind conditions for simulation runs}
\centering
\begin{tabular}{cccccc}
\hline
  & $N_{\rm p}~{\rm [cm^{-3}]}$ & $V~{\rm [km~s^{-1}]}$ & $B_{z,{\rm IMF}}~{\rm [nT]}$ & $M_{\rm f}$ & $\beta$  \\
\hline
normal solar wind & 4 & 400 & -3 & 5.3 & 5.0 \\
low-$\beta$ solar wind & 4 & 300 & -10 & 2.6 & 0.1 \\
\hline
\end{tabular}
\label{sw_params}
\end{table}

\section{Simulation Results}
\begin{figure}
\begin{center}
\includegraphics[width=0.75\textwidth]{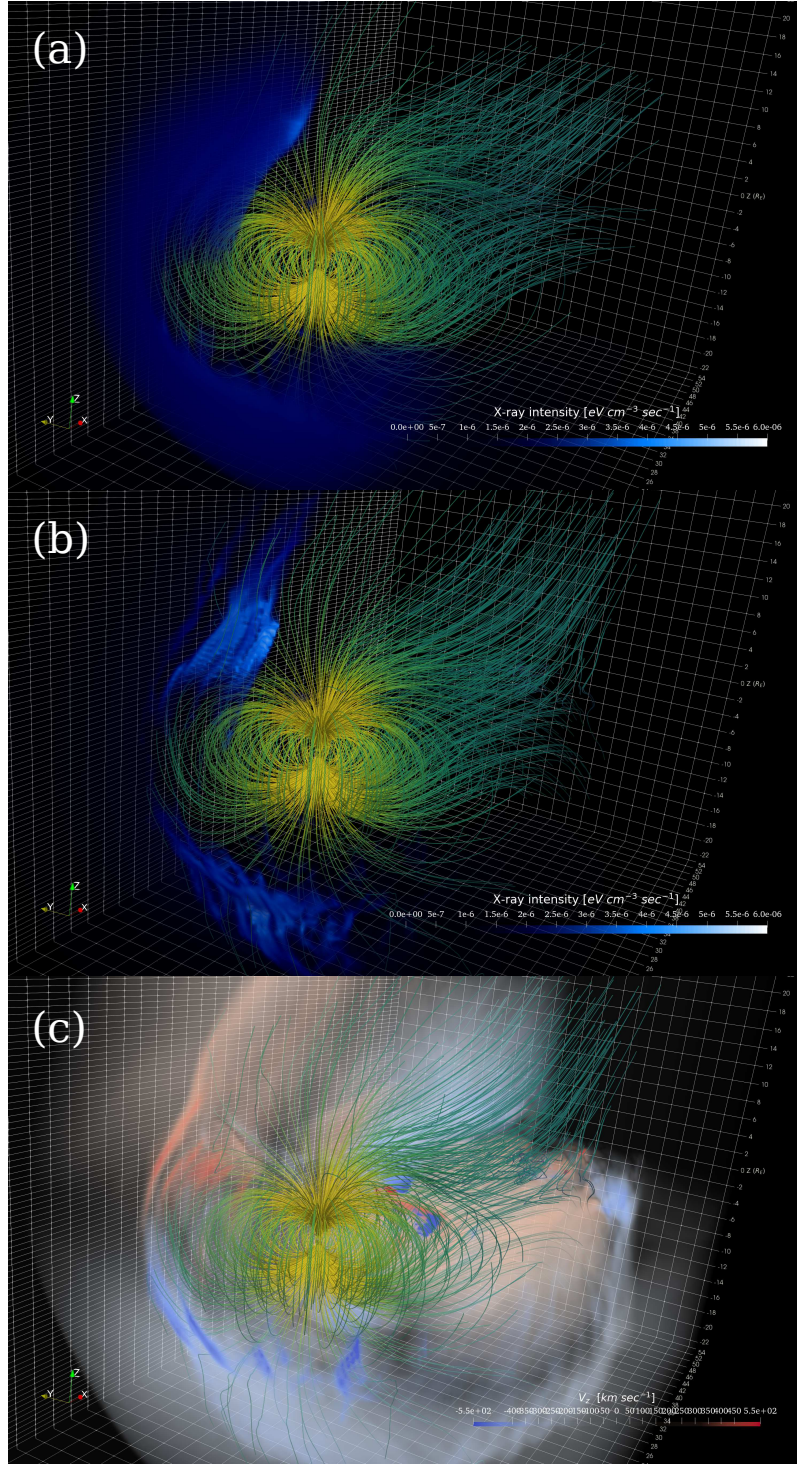}
\end{center}
    
\caption{3D distribution of the SWCX emission intensity for (a) the normal and (b) low-$\beta$ solar wind conditions along with (c) the $V_z$ profile for the low-$\beta$ condition. The color ranges with gradual opacity increment as indicated by the color bars. The color of the field lines represents the magnetic field strength. Data in the morning-north sector ($Y<0$ and $Z>0$) were dropped for 3D visualization.} 
\label{3D_view}
\end{figure}

We examined the global MHD simulations for the normal and low-$\beta$ solar wind conditions (Table \ref{sw_params}). We used data when the magnetosphere reached a stationary state after continuously injecting upstream plasma for an hour. 

Figure \ref{3D_view} shows global 3D images of the SWCX emission strength calculated by eq. (\ref{x-intensity}) for the normal (Figure \ref{3D_view}(a)) and low-$\beta$ (Figure \ref{3D_view}(b)) solar wind conditions. For the normal solar wind case, the X-ray emission is diffused in the entire dayside magnetosheath with strength of $\sim 2\times 10^{-6} {\rm\ eV}\ {\rm cm}^{-3}\ {\rm s}^{-1}$. A bright spot can be found at the top of the cusp region with strength of $\sim 3\times 10^{-6} {\rm\ eV}\ {\rm cm}^{-3}\ {\rm s}^{-1}$ where the dayside reconnection outflow meets the cusp region. By contrast, the emission intensity is remarkably brighter for the low-$\beta$ solar wind case. Some filamentary structures are found along the dayside magnetopause at different longitudinal locations with the strong emission strength of $\sim 6\times 10^{-6} {\rm\ eV}\ {\rm cm}^{-3}\ {\rm s}^{-1}$. These features indeed reflect plasma jets from the dayside reconnection regions, as shown in Figure \ref{3D_view}(c). The reconnection takes place locally in the azimuthal direction along the magnetopause, and produces very fast outflows reaching $600~{\rm km~s^{-1}}$ in both the northward (red) and southward (blue) directions.

\begin{figure}
\includegraphics[width=\textwidth]{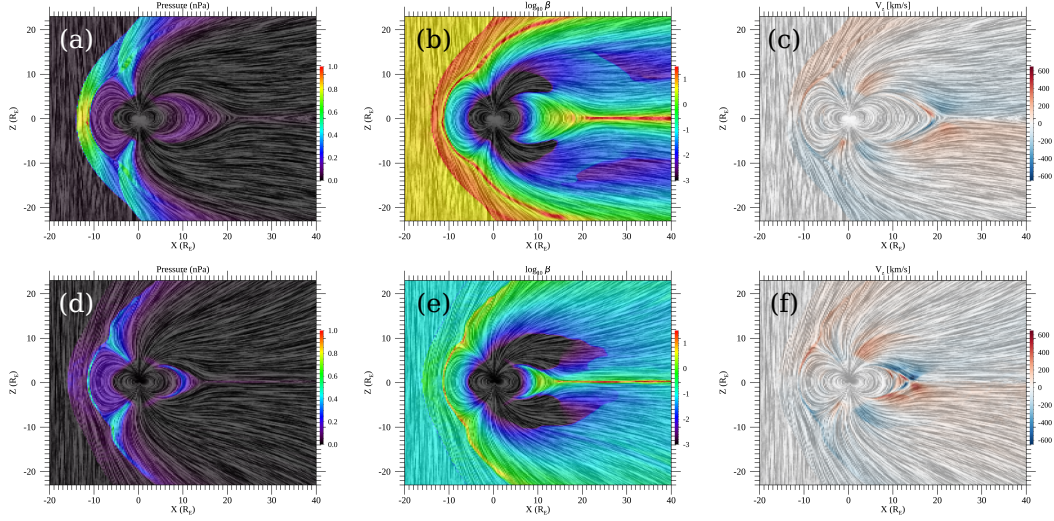}
\caption{Meridional profiles of the plasma pressure ((a) and (d)) in the unit of ($10^{-9}~{\rm Pa})$, plasma $\beta$ ((b) and (e)) in the logarithmic scale, and the z-component of the velocity ((c) and (f)) in ($\rm km~s^{-1}$) for the normal (top) and low-$\beta$ (bottom) solar wind conditions. Overplotted field lines represent the magnetic field.}
\label{2DPBV}
\end{figure}

Plasma and magnetic field profiles in the meridian plane are shown in Figure \ref{2DPBV}. Under the normal solar wind condition, the magnetosheath is essentially a high-$\beta$ plasma ($\beta > 10$) and the dayside magnetopause reconnection is rather moderate with an outflow speed of $\sim 200~{\rm km~s^{-1}}$. For the low-$\beta$ solar wind case, the pressure in the magnetosheath is as low as $0.2~{\rm nPa}$, and the resulting plasma $\beta$ in the magnetosheath $\beta_{\rm sheath} \sim 0.5$, as expected from the shock jump condition (Figures \ref{RH} and \ref{2DPBV}(e)). Because of the strong IMF, the reconnection outflow speed along the magnetopause reaches $|V_z| \sim 600~{\rm km~s^{-1}}$. Then, the flow decelerates by encountering the protruding cusp region where the plasma is adiabatically heated.

\begin{figure}
\includegraphics[width=\textwidth]{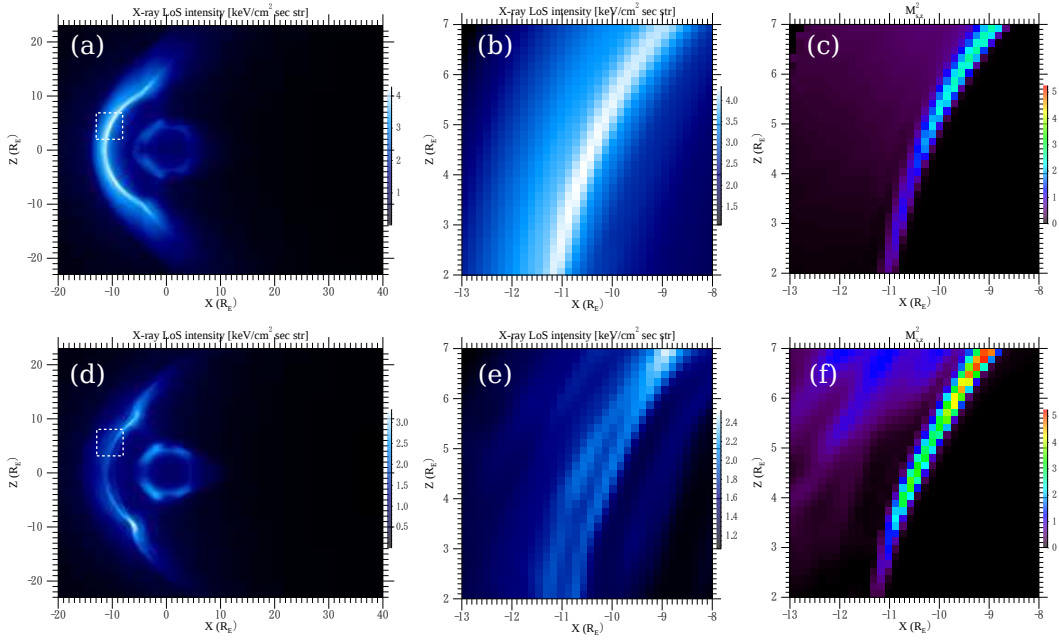}
\caption{Integrated X-ray intensity map ((a) and (d)), its enlarged view of the local magnetopause indicated by a square in (a) and (d) ((b) and (e)), and the sonic Mach number squared defined for the $V_z$ component in the meridian plane ((c) and (f)) for the normal (top) and low-$\beta$ (bottom) solar wind conditions. The bright region near the Earth reflects numerical artifacts around the inner boundary at $R=4~R_{\rm E}$}
\label{2DXM}
\end{figure}

The remote imaging technique provides a 2D map of the X-ray emission by integrating photons coming along the line of sight (LOS) of the telescope. Considering the low-latitude orbit of GEO-X and its distant orbit from the Earth ($R\sim 60~R_{\rm E}$), here we simply integrated the X-ray intensity along the Y-axis from a virtual observation location at $(X,~Y,~Z)=(0,~-60,~0)~R_{\rm E}$. We assumed zero emission outside of the simulation domain ($|Y|>38.3~R_{\rm E}$). Figure \ref{2DXM} shows the integrated X-ray intensity maps for the normal (Figure \ref{2DXM} (a)) and low-$\beta$ (Figure \ref{2DXM} (d)) cases along with enlarged views around the magnetopause (Figures \ref{2DXM}(b) and \ref{2DXM}(d)). As shown in previous studies, the integrated emission is strongest in the dayside magnetosheath (the bright region near the Earth reflects numerical artifacts around the inner boundary at $R=4~R_{\rm E}$). For the normal solar wind, the emission is bright in the entire magnetosheath region with a strength of $4.0~{\rm keV~cm^{-2}~s^{-1}~str^{-1}}$. This is quantitatively consistent with \citeA{connor_soft_2021} by taking into account their different solar wind number density ($10~{\rm cm^{-3}}$). Figure \ref{2DXM}(b) shows a 2D X-ray map in the localized $5~R_{\rm E} \times 5~R_{\rm E}$ area near the magnetopause. This area and the spatial resolution approximately correspond to the expected field-of-view and angular resolution of the GEO-X imager at a distance of $R=60~R_{\rm E}$, respectively \cite{ezoe_geo-x_2020}. With this area and spatial resolution, 2D X-ray maps would successfully identify the shape of the magnetosphere under a typical southward IMF condition. The strong emission was attributed mostly to the hot plasma both in the magnetosheath and in the current sheet, but the bulk motion of the reconnection outflow might have some contributions in the high-latitude region along the current sheet because the sonic Mach number gradually increases toward the cusp (Figure \ref{2DXM}(c)).

The integrated X-ray intensity is rather weak in the low-$\beta$ solar wind case (Figure \ref{2DXM}(d)) because the overall emission is weak and the strong emission area is localized (Figure \ref{3D_view}(b)). The magnetosheath is dark because the downstream plasma is still cold for the selected upstream condition. The emission strength is localized in the vicinity of the dayside magnetopause and the top of the cusp where the fast reconnection outflow decelerates and compresses the plasma (Figure \ref{2DPBV}(f)). When we focused on the localized area near the magnetopause, remarkably, two bright areas with $\sim 2.0~{\rm keV~cm^{-2}~s^{-1}~str^{-1}}$ were found (Figure \ref{2DXM}(e)), one of which coincides with a fast plasma jet in the meridian plane with large sonic Mach numbers $M_{\rm s}^2 > 2$ (Figure \ref{2DXM}(f)). Thus, the emission could be attributed to the bulk motion of the reconnection outflow (eq. (\ref{x-intensity_mach})). Another strong emission area also corresponds to a fast jet in a different longitudinal location, as indicated by the 3D profile of $V_z$ (Figure \ref{3D_view}(c)). These filamentary structures with large sonic Mach numbers imply the possibility of finding reconnection outflows from an actual LOS-integrated X-ray map particularly under low-$\beta$ solar wind conditions.

\section{Summary and Discussion}
The X-ray imager provides spectral information with emission lines in addition to a LOS-integrated 2D intensity map. A shift of the line energy and the line broadening provide information on bulk motion and thermal or turbulent motion of plasma \cite<e.g.,>{hitomi_collaboration_quiescent_2016}. Before using such spectral information, in this paper, we explored the possibility of extracting information on the bulk motion of plasma from an SWCX X-ray intensity map near the dayside magnetopause reconnection site by examining global MHD simulations. We found that under low plasma-$\beta$ solar wind conditions, the SWCX X-ray emission can reflect the reconnection outflow along the magnetopause. The emission from the reconnection outflows is somewhat faint ($\sim 2.0~{\rm keV~cm^{-2}~s^{-1}~str^{-1}}$) compared to the astrophysical backgrounds (e.g., $10.9~{\rm keV~cm^{-2}~s^{-1}~str^{-1}}$ from extragalactic sources \cite{cappelluti_chandra_2017}). However, these astrophysical sources can be considered constant within the dynamical time scales of the magnetospheric phenomena, and therefore, the SWCX emission can be obtained by subtracting the astrophysical origins from observation signals (Sibeck et al., 2018). More specifically, the dayside reconnection lasts up to several hours (Gosling et al., 1982; Phan et al., 2004), and the spatial extent of the structure is $\sim5\  R_{\rm E}$ along the outflow direction and a few thousands of kilometers ($\sim 0.5\ R_{\rm E}$) in the normal direction. These temporal and spatial scales can be resolved by the X-ray imager to be onboard GEO-X; the expected spatial (angular) resolution is $0.2\ R_{\rm E}$ from $60\ R_{\rm E}$ distance ($10\ {\rm arcmin}$), and the time cadence is within an hour (cf. Ezoe et al., 2020). We also note that there remains uncertainty in the predicted X-ray intensity by a factor of two depending on different exosphere models at these radial distances \cite{connor_soft_2021}.

Now, one may have a question about how we can actually observe such low-$\beta$ solar wind. Low-Mach-number or low-$\beta$ solar wind has been observed \cite{watari_statistical_2001,nishino_anomalous_2008,wilson_iii_statistical_2018}, in particular, associated with coronal mass ejections \cite{kataoka_flux_2006,lavraud_altered_2008}. Thus, although such solar winds are not usual, they are possible in some occurrence frequencies. For more quantitative discussion, we surveyed observation periods of the low-$\beta$ solar wind \add{under prolonged southward IMF,} satisfying the condition $M_{f1} \le 3.6$ and $\beta_1 \le 1.1$ (left-bottom area in Figure 1) \add{along with $B_{z,{\rm IMF}} \le -5\ {\rm nT}$} using the NASA OMNIWeb service. We found such particular solar winds can be found with \change{5.1}{1.0}\% probability during solar cycle 23 (August 1996--November 2008) and \change{3.0}{0.6}\% probability during solar cycle 24 (December 2008--November 2019). Thus, we can expect a certain observation feasibility in the upcoming solar cycle when the planned missions will be launched. In addition, plasma $\beta$ in the magnetosheath has been discussed in terms of dayside reconnection. \citeA{phan_dependence_2013} statistically studied the dayside reconnection occurrence frequency and found that the low-$\beta$ magnetosheath condition favored the magnetic reconnection occurrence as supported by the theories. \citeA{koga_dayside_2019} also examined the magnetosheath $\beta$ focusing on the reconnection outflow speed. There was a clear tendency that faster reconnection outflows were found under lower-$\beta$ magnetosheath conditions. Thus, the low-$\beta$ magnetosheath condition we proposed in this study has been positively supported by statistical studies using observational data.

Magnetic reconnection is essential for releasing the magnetic energy known as solar flares. There have been many opportunities to visually identify reconnecting field lines in the solar corona by solar observatories \cite<e.g.,>{yokoyama_clear_2001,liu_reconnecting_2010,savage_reconnection_2010,takasao_simultaneous_2012}. However, reconnection in the corona is complex in 3D and transient, and its behavior is difficult to understand. In this regard, dayside reconnection under low-$\beta$ solar wind conditions would provide great opportunities to visually understand the steady-state magnetic reconnection \cite{gosling_evidence_1982,phan_cluster_2004} and even transient phenomena as flux transfer events \cite<e.g.,>{akhavan-tafti_mms_2018} using SWCX X-ray imaging.

\section{Open Research}
The simulation data used in this study and the software (An Interactive Data Language script and ParaView state files) used to create the figures in this paper are available online \url{https://doi.org/10.5281/zenodo.6827016}.

\acknowledgments
We thank Masaki N. Nishino for the discussion on low-Mach-number solar wind studies. This work is supported by the Japan Society for the Promotion of Science (JSPS) KAKENHI Grant Number 20K20945. Numerical simulations were conducted using the FUJITSU Supercomputer PRIMEHPC FX1000 (Wisteria/BDEC-01) at the Information Technology Center, the University of Tokyo.

\bibliography{refs}

\end{document}